\documentclass[sigconf]{acmart}

\AtBeginDocument{%
  }

\setcopyright{acmlicensed}
\copyrightyear{2018}
\acmYear{2018}
\acmDOI{XXXXXXX.XXXXXXX}

\acmConference[Conference acronym 'XX]{Make sure to enter the correct
  conference title from your rights confirmation emai}{June 03--05,
  2018}{Woodstock, NY}
\acmISBN{978-1-4503-XXXX-X/18/06}

\usepackage{xspace}
\usepackage{balance}
\usepackage{makecell}
\usepackage{mathtools}
\usepackage{subfigure}
\usepackage{pgfplots}
\usepackage{adjustbox}
\usepackage{booktabs}
\usepackage{caption}
\usepackage{graphicx}
\usepackage{amsmath}
\usepackage{soul}
\usepackage{colortbl}
\usepackage{url}
\usepackage{footnote}
\usepackage{tikz}
\usepackage{array,multirow}
\usepackage{float}
\usepackage{enumitem}
\usepackage{enumitem,kantlipsum}
\usepackage{subcaption}
\usepackage{filecontents}
\pgfplotsset{compat=1.18}  
\usepackage{pgfplotstable}

\usepackage[most]{tcolorbox}
\tcbset{breakable, enhanced}

\makesavenoteenv{table}
\usepgfplotslibrary{groupplots}

\begin{document}

\newcommand{\ma}[1]{\textcolor{cyan}{Mahdi: #1}}
\newcommand{\system}{\textsc{Flowminator}\xspace}
\title{The Past Still Matters: A Temporally-Valid Data Discovery System}


\author{Mahdi Esmailoghli}
\affiliation{%
  \institution{Humboldt-Universität zu Berlin}
  \city{Berlin}
  \country{Germany}
}
\email{mahdi.esmailoghli@hu-berlin.de}

\author{Matthias Weidlich }
\affiliation{%
  \institution{Humboldt-Universität zu Berlin}
  \city{Berlin}
  \country{Germany}
}
\email{matthias.weidlich@hu-berlin.de}


\begin{abstract}
Over the past decade, the proliferation of public and enterprise data lakes has fueled intensive research into data discovery, aiming to identify the most relevant data from vast and complex corpora to support diverse user tasks. Significant progress has been made through the development of innovative index structures, similarity measures, and querying infrastructures. Despite these advances, a critical aspect remains overlooked: relevance is time-varying. Existing discovery methods largely ignore this temporal dimension, especially when explicit date/time metadata is missing. To fill this gap, we outline a vision for a data discovery system that incorporates the temporal dimension of data. Specifically, we define the problem of temporally-valid data discovery and argue that addressing it requires techniques for version discovery, temporal lineage inference, change log synthesis, and time-aware data discovery. We then present a system architecture to deliver these techniques, before we summarize research challenges and opportunities. As such, we lay the foundation for a new class of data discovery systems, transforming how we interact with evolving data lakes.
\end{abstract}

\maketitle

\section{Introduction}\label{sec:intro}
Over the past decade, a large number of public and private tabular data lakes have emerged, e.g., government open data~\cite{DBLP:conf/aaaiss/DingDGMLMH10, DBLP:journals/corr/abs-2106-09590}, enterprise-level lakes~\cite{DBLP:journals/pvldb/BharadwajGBG21,abs-2402-06282}, and crawled data lakes~\cite{Eberius:2015,DBLP:journals/pvldb/CafarellaHWWZ08, hulsebos2021gittables}.
The proliferation of data lakes has resulted in intensive research in data discovery, with the aim of identifying the most relevant data from these corpora to support diverse user tasks. 
Recent research has shown that identifying relevant data can drastically improve the performance of downstream tasks~\cite{DBLP:conf/edbt/EsmailoghliQA21, DBLP:journals/pvldb/ChepurkoMZFKK20}.
\begin{figure}
    \centering
    \includegraphics[scale=.16]{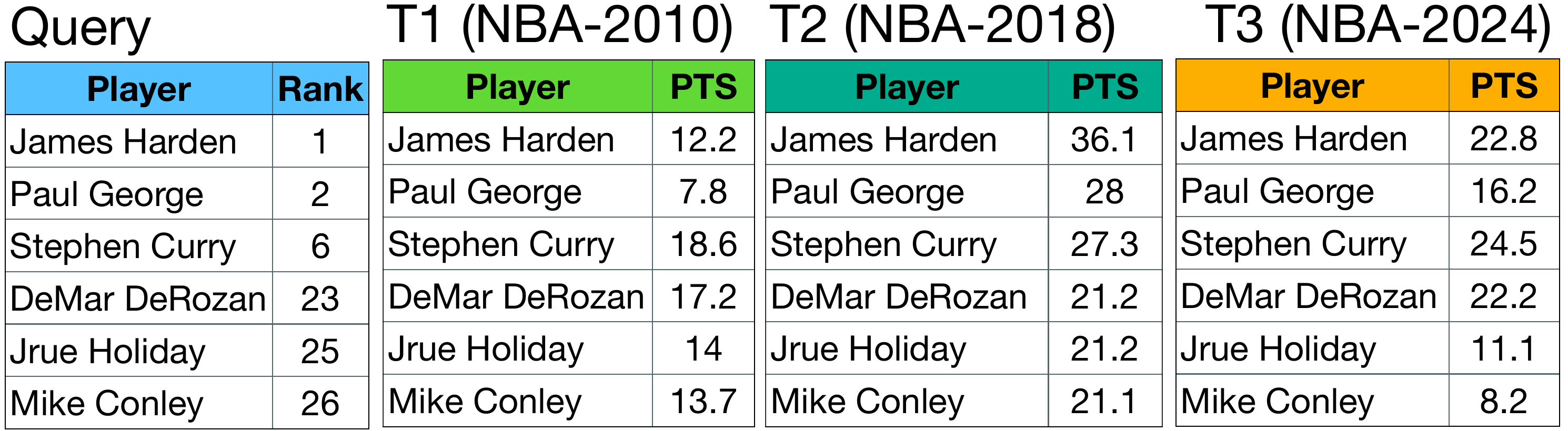}
    \vspace{-.9em}
    \caption{An example of a query and three joinable tables.}
    \label{fig:nba_example}
    \vspace{-1.5em}
\end{figure}

To enable efficient discovery of such data over large corpora, significant progress has been made through the development of innovative indexes, similarity measures, and end-to-end systems~\cite{esmailoghli2025blend}. 
Despite these advances, a critical aspect of data discovery remains overlooked: the temporal validity of discovered data.

Data lakes often contain multiple versions of datasets, accumulated over time~\cite{DBLP:journals/pvldb/ShragaM23}. 
While relevance is time-varying, current approaches treat each version of datasets individually, neglecting the fact that earlier versions of a dataset may no longer be valid for current analysis, or a time-sensitive task may require a specific version for reliable and reproducible results. 

\vspace{-.5em}
\begin{tcolorbox}[colback=white,boxrule=1pt, boxsep=0pt, sharp corners, left=5pt, right=5pt]
\emph{The gap between the current time-agnostic data discovery solutions and the temporal and semantic drifts in real-world data lakes results in ineffective, biased, and incorrect downstream analysis.} 
\end{tcolorbox}
\vspace{-.5em}

\noindent
The issues stemming from the neglect of temporal validity are manifold: ML model training becomes unreliable, statistical models derived from data encode incorrect information, and decision-making is based on semantically outdated information. Moreover, ignoring the evolution of data in discovery leads to an overwhelming number of similar results, which render it difficult for a user to separate the useful results from the outdated ones.

\vspace{.6em}
\noindent
\begin{tcolorbox}[colback=cyan!7, boxrule=0pt, boxsep=0pt, sharp corners, breakable,left=5pt, right=5pt,top=3pt,bottom=3pt]
\textit{Example.} Figure~\ref{fig:nba_example} shows four tables about NBA players: one query table and three candidate tables, all of which are joinable on the \textit{Player} column with the query table. 
The query table has a target column, \textit{Rank}, representing a player's rank in the league, while candidate tables contain points scored (\textit{PTS}).

Suppose the user aims to train a machine learning model to predict player \textit{Rank}. 
Although joining the query with any of the candidate tables highly boosts the predictive power of the model as it introduces a highly correlating feature \textit{PTS}, only joining with one candidate table results in a temporally-consistent final result.
In $2018$, ``James Harden'' ranked the best player in the NBA due to his outstanding performance, which is only observed in the corresponding candidate table. This makes the other two tables out-of-date for this specific task.
This example emphasizes 
how the semantic integrity of data can be compromised by ignoring the time drifts among candidate tables.
\end{tcolorbox}

In this vision paper, we unearth the problem of building a temporal data discovery system. We first discuss related work (\S\ref{sec:related}), before we define the problem of data discovery in \emph{temporal data lakes}, i.e., data lakes that contain multiple versions of tabular data (\S\ref{sec:problems}). We outline our vision for a system for \emph{temporally-valid data discovery}, i.e., for the identification of the correct table versions for a specific task (\S\ref{sec:system}). We elaborate on challenges and opportunities involved in building such a system (\S\ref{sec:challenges}), before we conclude the paper (\S\ref{sec:conclusion}).



\section{Related Work}\label{sec:related}
Temporal data discovery is relevant to several lines of research:
\subsection{Data Discovery}
Data discovery is the task of finding relevant data from a large corpus of tables. Discovery operators can be divided into five categories: keyword search~\cite{DBLP:conf/www/BrickleyBN19, DBLP:conf/cikm/ZhangSF21}, join discovery~\cite{DBLP:journals/pvldb/EsmailoghliQA22, DBLP:conf/sigmod/ZhuDNM19, DBLP:journals/pvldb/CasteloRSBCF21}, union discovery~\cite{DBLP:journals/pvldb/NargesianZPM18, DBLP:journals/pacmmod/KhatiwadaFSCGMR23, DBLP:journals/pvldb/FanWLZM23}, correlation discovery~\cite{DBLP:conf/icde/SantosBMF22, DBLP:conf/edbt/EsmailoghliQA21}, and exploratory data discovery~\cite{DBLP:conf/sigmod/NargesianPZBM20, DBLP:journals/pvldb/OuelletteSNBZPM21, DBLP:journals/tkde/NargesianPBZM23}.

Keyword search refers to finding data based on one or more keywords. These approaches either search among the metadata surrounding tables or tables including the provided keywords. 
Join discovery and union search approaches find tables that can extend the table at hand horizontally or vertically, respectively.
The input table for correlation discovery contains an additional target column. The goal is to find tables that not only are joinable to the input table on the given join key, but also contain a column that correlates with the target.
The exploratory search approaches aim at organizing tables based on semantic topics or overlaps for graph-based data explorations.
None of the aforementioned approaches accounts for temporality in the data by treating each table individually.

\subsection{Temporal Data Management}
A few papers consider data management in temporal data lakes.

Given two tables, Explain-Da-V~\cite{DBLP:journals/pvldb/ShragaM23} aims at discovering the changes that occurred in one table to generate the next. 
The authors propose different explanation methods based on column types. 
For instance, to explain the changes between two numerical columns, they train a regressor to obtain the coefficient and the intercept that transform one column into the other.
This approach assumes that the data versions are a priori known.
In addition, Explain-Da-V cannot explain semantic changes, similar to the example shown in Figure~\ref{fig:nba_example}.
Furthermore, the scalability concern of applying this approach to a large data lake remains unexplored.

Bornemann et al.~\cite{DBLP:conf/edbt/BornemannBKNNS24} introduce temporal inclusion dependency (tIND). The authors propose an index to find all tINDs for a given column. They demonstrate that confirming tINDs across versions is a strong indication of semantic inclusion dependency as opposed to a column subsuming the other only by chance in one version.

The dual streaming model~\cite{DBLP:conf/birte/SaxWWF18} addresses temporal data inconsistency in stream processing, particularly managing out-of-order data arrival. They model each operator as a sequence of updates to relational tables. Crucially, this model supports table versioning using timestamps, allowing older versions to be continuously and incrementally updated when out-of-order records arrive. 
Similar to Explain-Da-V, this approach makes strong assumptions that the data stream encapsulates the physical and logical order of the data, which is unavailable in the case of undocumented data lakes.

Bleifuss et al.~\cite{bleifuss2025schema} propose a schema change recommender for inherently evolving data lakes, e.g., Wikipedia.
The authors featurize past schema changes and generate rules to recommend how a table schema will be updated. 
Tables and their versions are already known through rich versioning capabilities in the Wikipedia history.
Each change is enriched with metadata, indicating fine-grain modifications and their orders in time. This information is missing in most of the less-controlled lakes. 

\subsection{Data Lake Systems}
Several systems have emerged to indirectly assist data discovery with a variety of constraints.
BLEND~\cite{esmailoghli2025blend} allows the user to generate discovery pipelines by connecting operators based on user-defined criteria. Although it implements a wide range of operators and their combinations, it does not directly support version-aware discovery.
Delta Lake~\cite{DBLP:journals/pvldb/ArmbrustDPXZ0YM20} is a storage that offers ACID properties over data lakes. It enables table versioning through an ordered identifier, making time travel and rollbacks possible. 
This is only useful if the tables are already annotated and the versions are already known. 
Moreover, although the availability of version identifiers facilitates time travel, it does not allow for temporal data discovery out of the box, and further discovery methods are needed to pinpoint the required data based on the user task.

\subsection{Traditional and Time Series Databases}
Time series databases (TSDBs), such as InfluxDB\footnote{\url{https://github.com/influxdata/influxdb}}, are optimized for querying data with explicit temporal attributes. 
Similarly, traditional databases support versioning through explicit transaction logs, assuming controlled environments, and complete metadata. However, these assumptions do not hold in data lakes, where versioning is implicit and metadata is missing. 
Our work addresses this gap by inferring version families, temporal lineage, and change operations from unlabeled tabular data, enabling temporally valid data discovery beyond the capabilities of TSDBs and conventional databases.
Nevertheless, techniques developed through decades of extensive research in building these systems can complement ideas and the vision we propose in this paper.
\section{Problem Characterization}\label{sec:problems}
\begin{figure}
    \center{\includegraphics[scale=0.14]
          {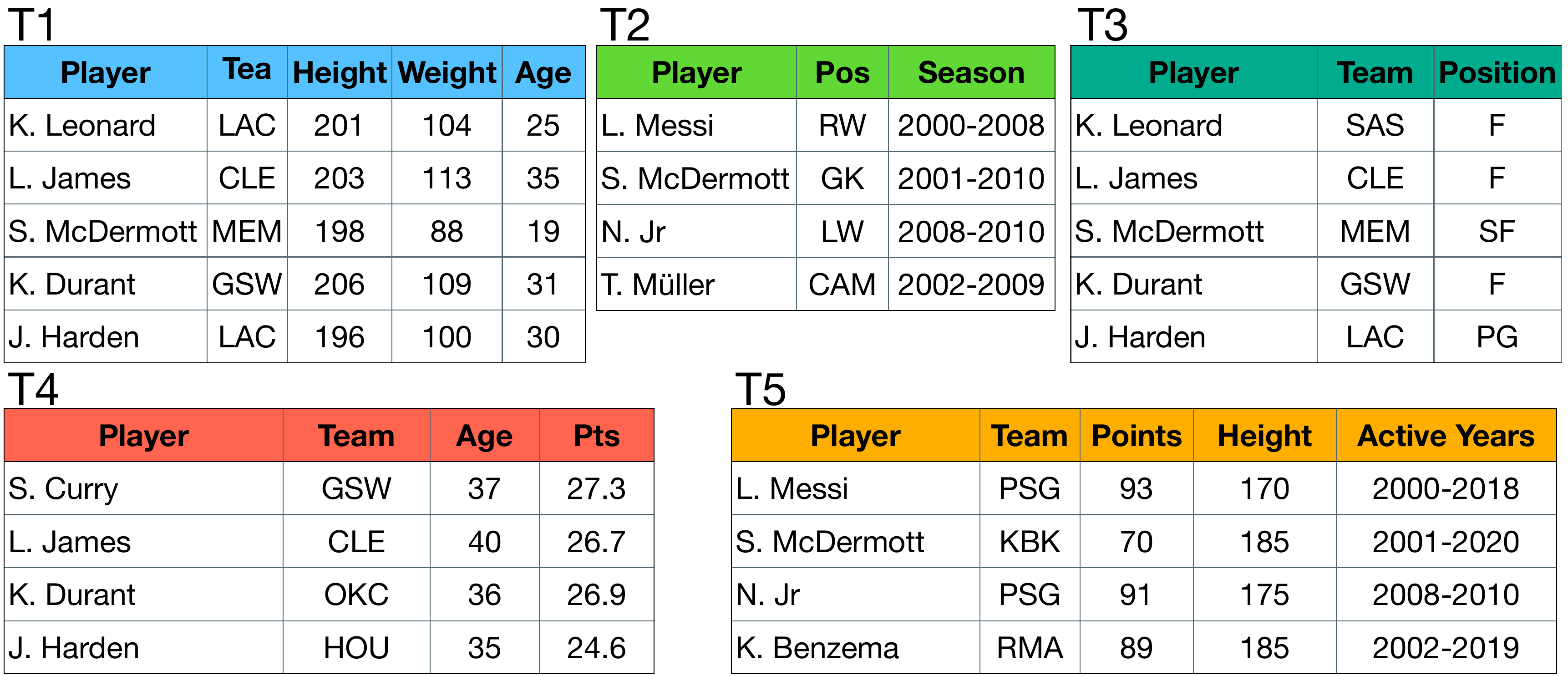}}
    \caption{Running example with two version families.}
    \vspace{-1em}
    \label{fig:running_example}
\end{figure}

In order to build an effective data discovery system for temporal data lakes, four major problems must be addressed: \textit{Version discovery}, \textit{Temporal lineage inference}, \textit{Change log synthesis}, and \textit{Temporally-valid data discovery}.
In this section, we explain these problems through a running example illustrated in Figure~\ref{fig:running_example}.

\subsection{Version Discovery}
Data lakes often lack metadata indicating dataset versioning. 
This issue can arise from a poorly storage solutions or a lake populated over many years. 
Temporal metadata issues also exists in corpora such as Wikipedia, which offers data versioning. 
Unlike Wiki pages that benefit from full change logging, table versions in Wiki pages do not necessarily receive the same versioning capabilities, resulting in inconsistencies between the text surrounding and explaining the table and the content of the table in a specific version of pages.
This lack of consistent metadata and versioning information makes the discovery of multi-version tables, which we call version families, a fundamental problem in data discovery. 

Figure~\ref{fig:running_example} shows two version families.
Tables T1, T3, and T4 form a version family, representing different versions of a dataset of NBA players. 
They contain attributes such as height, weight, team, position, age, and average points per game. 
Tables T2 and T5 represent versions of a dataset about FIFA players, containing columns such as player position, FIFA rating (points), and height.
While identifying version families in data discovery remains an underexplored research area, we argue that it presents significant challenges.

As illustrated by our running example, tables within the same version family can differ significantly in attributes, tuples, and semantics, making their identification non-trivial. 
For instance, tables T1 and T3 both describe NBA players but focus on different aspects: T1 includes height and weight, while T3 contains the player’s position. Neither table subsumes the other, highlighting the structural diversity within version families. Additionally, the team information differs due to temporal variations, as the tables were generated at different times when players belonged to different teams.

Another key challenge is that discovering different versions of the same dataset cannot rely on the available metadata alone. While many tables may share similar column names, such as Player, Team, and Points, their semantics can vary substantially. 
For example, in T4 (NBA), Points refers to the average points scored per game, e.g., 27.3, whereas in T5 (FIFA), it denotes a skill rating between $0$ and $100$, e.g., $93$ for L. Messi. Such inconsistencies in semantic meaning further complicate the discovery of version families.



Similarly, approaches based solely on table or attribute overlap can be misleading. Content-level semantics are essential, for instance, S. McDermott appears in both T1 (NBA) and T2 (FIFA) but refers to different individuals. Likewise, the Position column in T3 and T2 reflects domain-specific roles (F, SF, G in basketball versus RW, GK, LW in football), emphasizing the need for semantic understanding in version family discovery.

\subsection{Temporal Lineage Inference}
Once version families are discovered, it is necessary to extract the chronological order of tables within each family. This step is essential for temporally-aware data discovery, as the system must determine which data represents a newer version.

A common assumption is that newer versions contain more rows or columns. However, this is not always reliable, as tables within a family do not necessarily subsume one another. For instance, T2 includes a position attribute for players, while T4 does not, but instead provides the number of goals scored. Such structural variations highlight the complexity of establishing temporal order based solely on schema/table size.

Another solution to infer temporal order is to examine temporal indicators such as \textit{Date} or the progression of values in evolving attributes like \textit{Age}. For example, in T4, \textit{L. James} has an age of $40$, suggesting that this table is more recent than T1, in which he is $5$ years younger. Notably, this information is not constantly available.


\subsection{Change Log Synthesis}
To comprehend the lineage between data versions, it is essential to formalize how data evolves across version pairs. This evolution can be captured through fine-grained modification operations, collectively referred to as the change stream.

Explain-Da-V~\cite{DBLP:journals/pvldb/ShragaM23} attempts to generate such change explanations using data transformation operators. While relevant, this approach has several limitations. First, it relies on training regressors and classifiers, which do not scale to the size of modern data lakes.
Second, the model overlooks semantic drifts, where attribute meanings shift across versions (see Figure~\ref{fig:nba_example} for an example). 
Third, Explain-Da-V generates change streams for isolated table pairs, ignoring the fact that a single table may undergo multiple interdependent modifications before reaching a stable version. For instance, a user might update the content of a column, while another later corrects the column header to reflect the updated semantics.

The space of possible changes among versions is vast, domain-specific, and often not well-defined. Thus, it becomes necessary to define a minimal yet expressive set of change operators that can cover a broad range of version-to-version modifications. 
Designing such a set is inherently challenging.

Once these change operations are defined, the goal is to infer the sequence in which they were applied to transform an older version of a dataset into a newer one. Importantly, this change log should not be generated independently for every table pair, but rather treated as a unified and atomic representation of changes across the entire version family.

\subsection{Temporally-valid Data Discovery}

As a first step, it is essential to formalize what a temporal discovery query entails. How can a user specify the version of a dataset they need? As illustrated in Figure~\ref{fig:nba_example}, the appropriate version often depends on the specific analytical task. In some cases, users may already know the version they need, for example, requesting the latest version or a version that is newer or older than one they previously accessed. To support such diverse needs, we must design a query language that allows users to express temporal constraints in various forms, tailored to their tasks.

Once the expressiveness of the query is defined, an index structure can be built for fast discovery of relevant versions. However, traditional indexing techniques are not directly applicable here, since versioned datasets often lack explicit time attributes or intervals. Conventional temporal index structures, such as the T-index, interval trees, Multi-Version B-Trees (MVBT), TS-Index, and TP-index, assume well-defined temporal intervals or timestamps, which are typically absent or inconsistent in versioned data lakes.

Even with an efficient index structure data access remains a major bottleneck. Prior work in data discovery has shown that a significant portion of query time is spent on data retrieval from storage. Due to the immense size of data lakes, it is often infeasible to keep all data in main memory. Instead, datasets are stored in external databases, and accessing them on demand requires considerable overhead. Therefore, optimizing data access becomes a necessary component of enabling scalable and responsive temporally-valid data discovery. This is in particular crucial if accessing older table versions is prevalent and the system uses a multi-layer storage schema, e.g., Amazon S3 Intelligent-Tiering, which makes accessing older and infrequent data considerably slower.

\section{System Architecture}\label{sec:system}
\begin{figure*}
    \center{\includegraphics[scale=0.6]
          {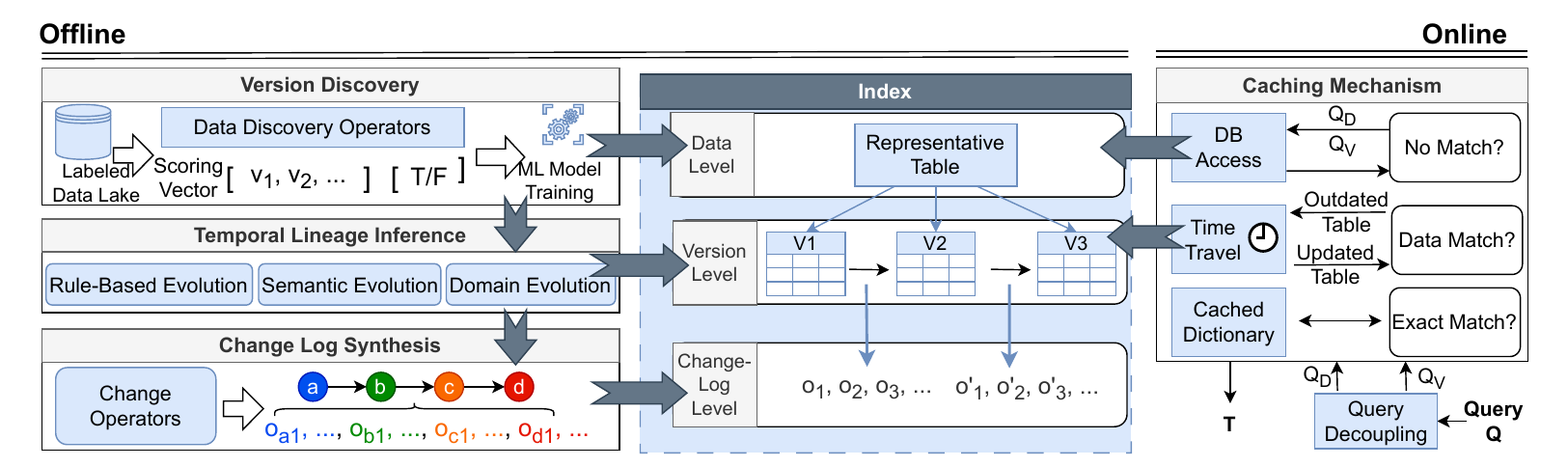}}
    \caption{System architecture.}
    \label{fig:architecture}
\end{figure*}

In this section, we describe our envisioned discovery system, which enables efficient indexing and querying of data lakes containing multi-version datasets.
Figure~\ref{fig:architecture} depicts the architecture of this system. It contains an offline phase, in which the data lake is ingested and the index structure is generated, and an online phase, where the index structure is used to find the most relevant tables (temporally and otherwise) to the given user query.

\subsection{Offline Phase}
This phase of the system is responsible for the index generation. 
The index should handle both traditional discovery queries, such as joins and unions, and temporally enriched queries that require identifying the appropriate version of relevant datasets.
To achieve this, we propose building a hierarchical index, in particular a three-level index structure: The top level allows for traditional index queries, the middle level serve as the storage for temporal lineage, in which different versions of datasets are stored, and in the bottom level, we index the change logs describing transformations from one version to another. We now explain the components involved in constructing this index.

\subsubsection{Version Family Discovery}\label{subsub:version_discovery}
To index versions of a table, the first step is identifying these version families, which are groups of tables that represent different versions of the same entity.
To achieve this, we propose an ML-based approach by leveraging a labeled data lake, such as Wikipedia tables with version histories. Using this labeled corpus, we can extract tables from different revisions of the same page and treat pairs of tables from the same page as positive examples (label True), and pairs from different pages as negative examples (label False). 
Then, one can train a binary classifier that predicts whether two tables are versions of the same entity.
To ensure a balanced training set, it is essential to include both intra-entity and inter-entity table pairs.

For feature extraction, we rely on a variety of data discovery techniques and their corresponding similarity metrics to generate a feature vector for each table pair. These features may include schema similarity, content similarity, containment similarity between column values, and context or semantic similarity scores. This multi-faceted feature set helps capture both structural, syntactic, and semantic signals relevant to version detection.

Once the features and labels are generated, we train the classifier to distinguish between table pairs that are different versions of the same entity and those that are not. Beyond classification, this approach also provides valuable insights into which features are most indicative of version relationships, highlighting characteristics that define version similarity.

After training, the model can be applied to unlabeled data lakes to discover version families, even in the absence of explicit version metadata, merely based on the differentiating similarity measures.

\subsubsection{Representative Table Generation}\label{subsub_representative_tables}
Current data discovery solutions are version agnostic, treating each table in the data lake as an independent entity. These systems typically compute a predefined similarity score for each table and return those with the highest scores in top-k search, or those exceeding a fixed threshold.
However, in data lakes containing temporally evolving datasets, users should be able to search for a dataset without being concerned about its evolution and available versions. 

To support temporally-valid discovery, the system must first find the queried entity then return the most relevant version of that entity. To achieve this, we propose generating a \textit{representative table} for each version family, a synthesized table that combines the information from all versions within the family.

The representative table is constructed using multi-table join and union operations, allowing it to capture the full range of content and schema elements present across versions. The goal is to ensure high similarity between query and the representative table iff user's desired table corresponds to any version within the family.

\subsubsection{Temporal Lineage Inference} \label{subsub:lineage_inference}
To materialize the concept of temporal evolution among tables in each version family, we must infer which table directly evolves from which, essentially identifying the immediate transformations between versions. 
Several heuristics can assist in this task.

One common heuristic is to use temporal columns containing date or time information. 
For instance, if the maximum date or time value in one table is later than those in others, this table likely represents the most recently updated version.

Another useful heuristic considers whether one table is a subset of another~\cite{DBLP:journals/pacmmod/PugnaloniZPLNS25, DBLP:conf/btw/KochEAA23}. If so, this suggests that the table has evolved vertically or horizontally by adding new rows or columns, respectively. This is a frequent pattern as new information becomes available and tables grow over time.

Beyond these rule-based approaches, it is important to evaluate the semantic or content evolution of tables. In particular, shifts in the distribution of data within certain columns can indicate real-world changes over time. For instance, people age, so the Age column in NBA player statistics naturally exhibits a predictable increase. Similarly, as shown in Figure~\ref{fig:nba_example}, player performance evolves smoothly over time as they approach their prime time ($2018$ in the example) or retirement ($2024$ in the example).
Tracking these distributional changes across tables can help reveal the correct order of versions within a version family.

Finally, large language models (LLMs) provide a powerful resource with broad domain knowledge. LLMs can serve as a tool when more direct heuristics fail, especially in cases where the data is domain-specific or when it lacks explicit temporal information.

\subsubsection{Change Log Synthesis} \label{subsub:change_log}
Once version families are discovered and the datasets within each family are ordered based on their conceptual evolution, it becomes possible to generate the transformations that produced each version. Explain-Da-V generates such explanations by leveraging the known order of evolution between pairs of tables~\cite{DBLP:journals/pvldb/ShragaM23}. 
However, two main challenges arise when discovering these change logs.

First, there is the issue of scalability. Change logs must be generated for every pair of tables within each version family across the entire data lake. Although this process can be performed offline, it remains computationally expensive because the complexity of the Explain-Da-V approach is relatively high.

Second, version families often contain more than just two versions. We argue that change logs should be generated with consideration of the entire evolution history, rather than focusing solely on isolated pairs. This is because these changes are not necessarily independent, as multiple versions of the same table might be created as part of a single update.

\subsubsection{Index construction}
Once version families are discovered, representative tables are generated, the temporal lineage of versions is inferred, and change logs are obtained, the index structure can be constructed for efficient retrievals.

The top level of the index, called the Data Level, stored the representative tables. The storage format may vary depending on the expected data discovery tasks and methods employed. We recommend using the index proposed in BLEND~\cite{DBLP:journals/corr/abs-2310-02656}, as it supports a broad range of data discovery tasks.

At the Version level, the ordered versions of each representative table are stored. As the number of these tables per representative table is expected to be limited, one can store them as linked lists.

In the change log level, change logs will be indexed based on their corresponding version table. It is the table that, by applying the given sequence of change logs, one can generate the next consecutive table in the Version level.
This level can be stored as B+ tree, where the leaves are the operators, allowing us to move from one operator to the next, plus an indicator where the current change log ends and the new one starts, enabling us to apply a specific number of change logs on the data and generate multiple versions.

\subsection{Online Phase}
In the online phase, the user provides the query $Q$ and the system looks up the index and outputs the correct data version $T$.
The online phase is comprised of query decoupling and version lookup.

\subsubsection{Query Decoupling}\label{subsubs:decoupling}
We split the data lake to representative tables and their corresponding versions.
Therefore, the system analyzes the query and the user task to extract two sub-queries, namely the dataset query ($Q_D$) and the version query ($Q_V$). 
$Q_D$ is used to obtain the relevant representative table and $Q_V$ is used to obtain the temporally-valid version among all candidates.
Query decoupling can be conducted in two way: explicit and implicit.
Explicit decoupling is done based on the clear instructions provided by the user. For instance, the user can demand the most recent version of the matching dataset. In this case, the system retrieves only the final and most up-to-date version of the found dataset. 
Implicit decoupling requires further processing of the input. For instance, as depicted in Figure~\ref{fig:nba_example}, likely, the user does not recognize the best version of the dataset.
In this case, the system needs to generate a version query based on implicit information encapsulated in the discovery query. 
In particular, the version query in our example indicates that the \textit{PTS} for \textit{James Harden} must be maximum because he earned the first rank during that season. 

\subsubsection{Table Version Lookup}
Building the index, the system can answer temporally-valid queries, however, for further efficiency, we propose a cache-based lookup strategy. It operates based on the match of the input queries with the target table.
Given the input query components, i.e., $Q_D$ and $Q_V$, if both queries hit a cached data version in the in-memory dictionary, the table is fetched and returned to the user.

If only $Q_D$ is matched but the version is not available, instead of conducting a full database search, which is inherently slow, we propose to rely on time travel. The time travel operator fetches an already cached table as well as the change log operators that can be located fast in the index based on the table identifier, or can be read from memory, as change logs are expected to be smaller than storing the whole version family. 
Then, using the available version and the fetched change log, the time travel operator constructs the desired table on-the-fly and returns it as the query result.

In the case where neither the data nor the version is matched, the system accesses the index structure in the database to fetch the table and then caches it for future queries.

\section{Challenges and Opportunities}\label{sec:challenges}
To build this system, several challenges must be addressed:

\vspace{.1em}
\noindent\textbf{Version variability.}
In Section~\ref{subsub:version_discovery}, we discussed training a classifier that predicts whether a pair of tables belong to the same version family. While training a model based on common data discovery measures can provide valuable insight in similarities that define version families, achieving this is a challenging task. 
This is because tables in the same version family can be arbitrarily different. Intuitively, two versions that are the result of a single modification, e.g., adding a new column, are more similar compared to two versions that are the results of years of modifications and updates.
Due to this complexity, one might need to explore complex feature spaces and obtain a deep knowledge of discovering these version families.

\vspace{.1em}
\noindent\textbf{Version integration.}
Section~\ref{subsub_representative_tables} introduces representative tables. Although all-in-one tables simplify the discovery by separating data and version lookups, constructing them remains a challenging problem~\cite{DBLP:journals/pvldb/KhatiwadaSGM22}.
Tables of a version family can differ in various dimensions, thus, their aggregation into one representative table cannot be guaranteed. For instance, the schema of tables might not be compatible, leading to issues in unioning tables, or tables may not highly overlap in columns, limiting the ability to horizontally merge them. 
Furthermore, naively integrating multiple tables into one might result in a sparse result, making it challenging to efficiently store or query. It is also necessary to investigate how different integration approaches can impact the results of the data discovery with respect to false positives and false negatives. 

\vspace{.1em}
\noindent\textbf{Non-linear evolution.}
In Sections~\ref{subsub:lineage_inference} and \ref{subsub:change_log}, we elaborate on discovering temporal lineage as well as change logs in version families. For simplicity, we limited the discussion to linear data evolution, where a new version is always obtained based on the latest table.
However, data evolution is not necessarily linear. 
One version might not be generated from the latest version, but rather from an older one. This not only results in a DAG-shaped lineage but also can lead to redundant and even duplicate instances, making the data discovery, as well as version discovery, even more challenging.

\vspace{.1em}
\noindent\textbf{Reverse time travel.}
The state of the art~\cite{DBLP:journals/pvldb/ShragaM23} only allows for time travel in one direction. The system should also reverse time travel. This means that the system should be able to rebuild the older version of a given table. To this end, one should generate the list of inverse operators used to generate the next version. Reverse time travel brings the system an ultimate flexibility and storage efficiency, requiring it to save only a limited number of versions.

\vspace{.1em}
\noindent\textbf{Version query extraction.}
We discussed the idea of decoupling data and version queries (Section~\ref{subsubs:decoupling}. 
Although this simplifies the discovery of the correct table, it is not a straightforward task.
In particular, implicit decoupling remains a challenge, where the user cannot explicitly express the desired version. It requires a deep understanding of the semantic relations between data discovery tasks and the versions available in the lake.
\section{Conclusion}\label{sec:conclusion}
This paper addresses data discovery under temporal drift, focusing on the challenge of finding multiple temporal versions of the same dataset, which are often treated independently by existing methods.

We propose a two-phase architecture: an offline phase that builds a temporally-enriched index, and an online phase that leverages it for efficient and temporally-valid retrieval. 
Our classifier distinguishes unrelated datasets from temporally linked versions.
We introduce heuristics for constructing version lineages and extracting change logs to support temporal navigation. To improve query handling, we decouple content-based and version-specific queries, inferring both from user input.
Finally, we present a hybrid caching strategy that combines in-memory storage with on-demand time travel, reducing latency and improving dataset discovery at scale.

\balance

\small
\bibliographystyle{abbrv}
\bibliography{references} 

\end{document}